\documentclass[lettersize,journal]{IEEEtran}
\usepackage{amsmath,amsfonts}
\usepackage{algorithmic}
\usepackage{algorithm}
\usepackage{array}
\usepackage[caption=false,font=footnotesize,labelfont=rm,textfont=rm]{subfig}
\usepackage{textcomp}
\usepackage{stfloats}
\usepackage{url}
\usepackage{verbatim}
\usepackage{graphicx}
\usepackage{cite}
\usepackage{xcolor}
\newcommand{\rev}{\textcolor {blue}}
\newcommand{\que}{\textcolor {red}}
\newcommand{\yx}{\textcolor {teal}}

\begin{document}

\title{Sensing Aided Uplink Transmission in OTFS ISAC with Joint Parameter Association, Channel Estimation and Signal Detection} 

\author{Xi Yang, Hang Li, Qinghua Guo,~\IEEEmembership{Senior Member, IEEE}, J. Andrew Zhang,~\IEEEmembership{Senior Member, IEEE}, \\Xiaojing Huang,~\IEEEmembership{Senior Member, IEEE}, and Zhiqun Cheng,~\IEEEmembership{Member, IEEE}.
        % <-this % stops a space
        %\thanks{This work is supported by National Key R\&D Program of China (Grant:2018YFE0207500), National Natural Science Foundation (Grant 62071163), Zhejiang Provincial Natural Science Foundation (Grant LY22F010003) and Project of Ministry of Science and Technology (Grant D20011). 
        %\par Xi Yang, Hang Li and Zhiqun Cheng are with the School of Electronics and Information, Hangzhou Dianzi University, Hangzhou 310018, China (e-mail: yancy486@163.com; hangli@hdu.edu.cn; zhiqun@hdu.edu.cn). 
        %\par Qinghua Guo is with the School of Electrical, Computer and Telecommunications Engineering, University of Wollongong, Wollongong, NSW 2522, Australia (e-mail: qguo@uow.edu.au). 
        %\par J. Andrew Zhang, Xiaojing Huang are with the Global Big Data Technologies Centre, University of Technology Sydney, Ultimo, NSW 2007, Australia (e-mail: andrew.zhang@uts.edu.au; Xiaojing.Huang@uts.edu.au).}% <-this % stops a space
        \thanks{Xi Yang, Hang Li and Zhiqun Cheng are with the School of Electronics and Information, Hangzhou Dianzi University, Hangzhou 310018, China (e-mail: yancy486@163.com; hangli@hdu.edu.cn; zhiqun@hdu.edu.cn). 
        \par Qinghua Guo is with the School of Electrical, Computer and Telecommunications Engineering, University of Wollongong, Wollongong, NSW 2522, Australia (e-mail: qguo@uow.edu.au). 
        \par J. Andrew Zhang and Xiaojing Huang are with the Global Big Data Technologies Centre, University of Technology Sydney, Ultimo, NSW 2007, Australia (e-mail: andrew.zhang@uts.edu.au; Xiaojing.Huang@uts.edu.au).}

        }

        \markboth{}%
        {}

        \maketitle

\begin{abstract}
In this work, we study sensing-aided uplink transmission in an integrated sensing and communication (ISAC) vehicular network with the use of orthogonal time frequency space (OTFS) modulation. 
To exploit sensing parameters for improving uplink communications, the parameters must be first associated with the transmitters, which is a challenging task. 
We propose a scheme that jointly conducts parameter association, channel estimation and signal detection by formulating it as a constrained bilinear recovery problem. Then we develop a message passing algorithm to solve the problem, leveraging the bilinear unitary approximate message passing (Bi-UAMP) algorithm. 
Numerical results validate the proposed scheme, which show that relevant performance bounds can be closely approached.
\end{abstract}

%\vspace{-0.3cm}
\section{Introduction}
%\IEEEPARstart{R}{ecently}, %the development of 5G network and even the realization of 6G network in the future, 
Recently, the integrated sensing and communications (ISAC) based intelligent transportation has received tremendous attention, as it can potentially enable various applications such as autonomous driving, traffic management, and Internet of Vehicles (IoV) \cite{ref1,ref2}.
%It can not only realize radar sensing and wireless communication, but also mutually improve the performance of two functions.
Sensing the states (e.g., locations and speeds) of vehicles and surrounding objects improves the safety of IoV effectively, as well as the reliability of communications aided by sensing. 
In order to meet the needs of high-mobility, orthogonal time frequency space (OTFS) modulation has been employed in ISAC systems \cite{ref3,ref4,ref5,ref6}. In this work, we focus on OTFS uplink transmission in an ISAC vehicular network.
%OTFS is a promising technology to overcome the inter-carrier interference of traditional orthogonal frequency division multiplexing (OFDM) signals against high mobility by modulating symbols in the delay-Doppler (DD) domain\cite{ref6}.
\par
%\vspace{0.2cm}
Various OTFS signal detectors have been proposed in the literature, e.g., the message passing based detectors \cite{ref7,ref8,ref9}.
%For example, efficient message passing (MP) equalization algorithms were proposed in \cite{ref7} for signal detection with a fractionally spaced sampling. A signal detector with high performance while low complexity was proposed in \cite{ref8} based on the unitary approximate message passing (UAMP) \cite{ref15}.} 
These detectors require accurate channel state information, which can be acquired using different ways.   
%In order to improve the robustness of signal detection, Yuan $et$ $al$. designed an OTFS receiver with low complexity and excellent performance using the unitary approximate message passing (UAMP) algorithm, which however generally requires accurate channel state information (CSI)\cite{ref8}.
%The acquisition of CSI is mainly through two ways.
The channel can be estimated before data transmission \cite{ref7,ref8,ref9}, which occupies channel coherence time and leads to substantial overhead. 
One can also separate the pilot symbols and data symbols in the delay-Doppler (DD) domain using the guard intervals \cite{ref10}, which, however, results in large overhead.
To avoid the spectral loss, pilot symbols can also be superimposed with data symbols, where the interference between data and pilot needs to be carefully handled with joint channel estimation and signal detection \cite{ref11}. 
In \cite{ref12}, sensing-aided communications in IoV was proposed to reduce the delay and overhead, and improve the communication performance. However, the parameter association issue is not considered.

\par
In this work, we propose a novel scheme for uplink transmission with sensing-assisted joint channel estimation and signal detection in an ISAC OTFS vehicular network. 
To achieve this, the sensing parameters acquired by the roadside unit (RSU) through downlink sensing need to be first associated with the vehicles in the network. 
To tackle this challenging problem, we propose joint parameter association, channel estimation and signal detection (PACESD). 
We show that the joint PACESD can be formulated as a constrained bilinear recovery problem, where parameter association and channel estimation lead to a sparse vector that needs to be jointly recovered with a discrete-valued symbol vector. 
Then, leveraging the bilinear UAMP (Bi-UAMP) algorithm \cite{ref13}, we develop a message passing based Bayesian algorithm to solve this problem.    
Numerical results validate the proposed scheme and show that the proposed algorithm can achieve performance close to the relevant bounds.

\par 
\emph{Notations}: Unless otherwise specified, we use a boldface lowercase letter, a boldface capital letter, and a calligraphy letter to denote a vector, a matrix, and a set, respectively;
the $\mathbb{C}$ denotes the complex number field;
$\mathrm{vec}(\cdot)$ and $\otimes$ denote the vectorization and the Kronecker product operator, respectively;
the superscripts $(\cdot)^{T}$, $(\cdot)^{H}$ and $(\cdot)^{*}$ denote the transpose, the conjugate transpose and conjugate operation\yx{s}, respectively;
$\mathbf{F}_{N}$ and $\mathbf{I}_{N}$ denote $N$-dimensional discrete Fourier transform (DFT) matrix and identity matrix, respectively;
$\mathbf{1}$ and $\mathbf{0}$ denote the all-ones vector and all-zeros vector, respectively;
$\delta \left ( \cdot  \right )$ denotes the Dirac delta function;
$f(x)\propto g(x)$ denotes the relation $f(x)=cg(x)$ for some positive constant $c$;
$|\cdot|^2$  denotes the element-wise magnitude squared operation;
$\|\cdot\|$ denotes the $l_2$ norm;
$\mathbf{a}\cdot \mathbf{b}$ and $\mathbf{a}./\mathbf{b}$ denote the element-wise product and division of the two vectors, respectively;
$<f(\mathbf{x})>_{g(\mathbf{x})}$ denotes the expectation of $f(\mathbf{x})$ with respect to probability density function $g(\mathbf{x})$;
$\mathcal{M}_{a\rightarrow b}(x)$ denotes a message passed from node $a$ to $b$ ,which is a function of $x$;
$b(x)$ denotes the belief of $x$.

%\vspace{-1.55cm}

\begin{figure}[tt]
  \setlength{\abovecaptionskip}{0.cm}
  \setlength{\belowcaptionskip}{-0.cm}
    \centering
    \vspace{-0.4cm}
    \includegraphics[width=3in]{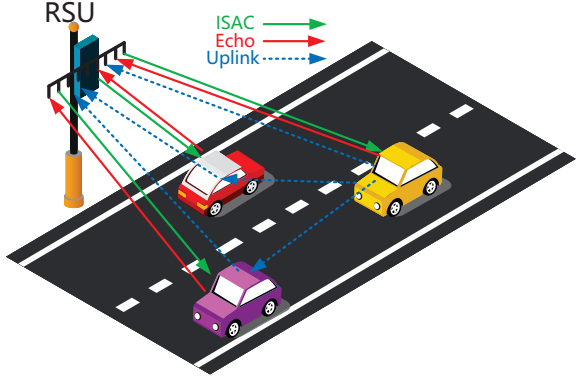}
    \caption{Illustration of an ISAC vehicle network.}
    \label{fig_1}
    \vspace{-0.3cm}
  \end{figure}
  \vspace{-0.3cm}
\section{ISAC Vehicular Network and OTFS Modulation}
\vspace{-0.1cm}
\subsection {ISAC Vehicular Network}

As shown in Fig. \ref{fig_1}, we consider an ISAC vehicular network, where the RSU provides communication services to multiple vehicles, while sensing the targets in the area. Assume that the RSU is equipped with a uniform linear array with $N_{BS}$ antennas and each vehicle has a single antenna. We focus on uplink transmission, i.e., vehicles transmit signals to the RSU, which is assisted by the sensing function of the network.\footnote{Although pilots can be used to estimate channel state information (CSI) for signal detection, sensing is still necessary for dynamic network monitoring and vehicle safety improvement.}
The workflow is as follows. 
\subsubsection{State estimation}
The RSU first sends broadcast signals, which are reflected by the targets/vehicles within the service area and received by the RSU. 
With the received echo signals, the RSU estimates the sensing parameters including the arrival angles, time delays, and Doppler frequencies, based on which the RSU can determine the locations and speeds of the targets.
%\rev{It is noted that the acquisition of the sensing parameters has been investigated in the literature, e.g., the works in [?] and [?].}
%The RSU sends broadcast signals to targets (vehicles and non-communication objects) within the service range. The reflected echo signals are received by RSU for estimating the sensing parameters of delay, Doppler, and angle values of all targets relative to the RSU. 
\subsubsection{State prediction}
With the estimated parameters, the RSU predicts the states of targets/vehicles in the next timeslot for uplink communication.
\par
\subsubsection{Sensing aided uplink transmission and parameter association}
Then, with time division multiple access (TDMA), the vehicles transmit communication signals to the RSU \cite{ref12}, and the RSU performs channel estimation and signal detection with the aid of sensing parameters acquired in Step 2. Parameter association is jointly conducted in this process.
\par
It is noted that sensing parameter acquisition and prediction in Steps 1 and 2 have been investigated in the literature, e.g., in \cite{ref5,ref12}.
This work focuses on Step 3 for joint PACESD. 
%\vspace{-0.8cm}
\subsection{OTFS Modulation}

%OTFS modulation is employed by the network.
Let $\mathbf{X}\in \mathbb{C}^{M\times N}$ denote the transmitted data symbol matrix in the DD domain, where $M$ and $N$ denote the numbers of subcarriers and time slots, respectively. 
Let $\Delta f$ and $T$ be the subcarrier spacing and time slot duration, respectively, where $T\Delta f=1$. 
Then, the transmitted signal $\mathbf{S}\in \mathbb{C}^{M\times N}$ in the time-delay (TD) domain is given by \cite{ref14}
\begin{equation}
    \label{eq_1}
    \mathbf{S}=\underset{\mathrm{Heisenberg\,transform}}{\underbrace{\left(\mathbf{G}_{\mathrm{tx}}\mathbf{F}_{M}^{H}\right)}}\times\underset{\mathrm{ISFFT}}{\underbrace{\left(\mathbf{F}_{M}\mathbf{X}\mathbf{F}_{N}^{H}\right)}},
\end{equation}
where the inverse symplectic finite Fourier transform (ISFFT) is equivalent to an $M$-point FFT of the columns and an $N$-ponit IFFT of the rows of $\mathbf{X}$, 
$\mathbf{G}_{\mathrm{tx}}=$$\mathrm{diag}[g_{\mathrm{tx}}(0) ,g_{\mathrm{tx}}(T/M) ,\dots,g_{\mathrm{tx}}((M-1)T/M)]\in \mathbb{C}^{M\times M}$, and $g_{\mathrm{tx}}(t)$ is the pulse-shaping waveform.
In this work, we employ the rectangular waveform, i.e., $\mathbf{G}_{\mathrm{tx}}$ is an identity matrix $\mathbf{I}_{M}$.
The transmitted signal vector $\mathbf{s}\in \mathbb{C}^{MN\times 1}$ can be represented by vectorizing the TD domain signal matrix $\mathbf{S}$, i.e.,
\begin{equation} 
    \begin{aligned}
    \label{eq_2}
    \mathbf{s}=\mathrm{vec}\left(\mathbf{I}_{M}\mathbf{X}\mathbf{F}_{\mathrm{N}}^{H}\right)=\left(\mathbf{F}_{\mathrm{N}}^{H}\otimes \mathbf{I}_{M}\right)\mathbf{x},
    \end{aligned}    
\end{equation}
where $ \mathbf{x}= \mathrm{vec}\left(\mathbf{X}\right)$.

%\subsection{Uplink Transmission}

Assuming that the number of independent resolvable paths between the RSU and Vehicle $i$ is $P_{i}$, the TD domain channel can be represented as \cite{ref12}
\begin{equation}
  \label{eq_3}
  \mathbf{h}_{i}\left ( \tau,\nu \right )=\sum\nolimits_{p=1}^{P_{i}}h_{i,p}\mathbf{b}\left ( \theta_{i,p} \right )\delta \left ( \tau-\tau_{i,p} \right )\delta \left ( \nu-\nu_{i,p} \right ),
\end{equation}
where $h_{i,p}$, $\tau_{i,p}$, $\nu_{i,p}$ and $\theta_{i,p}$ denote the channel gain, the delay, the Doppler frequency, and the angle relative to the RSU for the $p$-th path of vehicle $i$, respectively, and
$\mathbf{b}\left ( \theta_{i,p} \right )=1/\sqrt{N_{\mathrm{BS}}}[1,e^{j\pi \mathrm{sin}\theta _{i,p}},\cdots,e^{j\pi (N_{\mathrm{BS}}-1)\mathrm{sin}\theta _{i,p}}]^{T}$ is the receive steering vector. Assuming a cyclic prefix is used in each OTFS block, we define the channel matrix \cite{ref15} %\eqref{eq_3}
\begin{equation}
  \label{eq_4}
  \mathbf{H}_{i}=\sum\nolimits_{p=1}^{P_{i}}h_{i,p}\mathbf{b}\left ( \theta _{i,p} \right )\mathbf{\Pi}^{l_{i,p}}\mathbf{\Delta}^{k_{i,p}},
\end{equation} 
where the permutation matrix $\mathbf{\Pi}$ is obtained by shifting the first column of an identity matrix to the last column (see (9) in \cite{ref15}). 
$\mathbf{\Delta}=\mathrm{diag}\left \{ 1, e^{\frac{j2\pi }{MN}},\dots,e^{\frac{j2\pi\left ( MN-1 \right ) }{MN}}\right \}$ is a diagonal matrix characterizing the Doppler influence,
$l_{i,p}=\tau_{i,p}M\Delta f$, $0\leq l_{i,p}\leq M-1$ and $k_{i,p}=\nu_{i,p}NT$, $0\leq k_{i,p}\leq N-1$ denote the indices of delay and Doppler associated with the $p$-th path of vehicle $i$, respectively.
\par
The RSU uses a bank of receive beamformers $\mathbf{f}_{p}\in \mathbb{C}^{N_{BS}\times 1}$ to receive the signals transmitted by vehicle $i$, and the received signal in the TD domain can be expressed as \cite{ref12}
\begin{equation}
  \label{eq_6}
  \mathbf{r}_{i}=\sum\nolimits_{p=1}^{P_{i}}h_{i,p}\left(\mathbf{f}_{p}^{H}\mathbf{b}\left ( \theta _{i,p} \right )\right)\mathbf{\Pi}^{l_{i,p}}\mathbf{\Delta}^{k_{i,p}}\mathbf{s}_{i}+\mathbf{q}_{i},
\end{equation}
where $\mathbf{q}_{i}$ denotes the additive white Gaussian noise (AWGN) vector in the TD domain.
With \eqref{eq_2} and \eqref{eq_6}, the received signal in the DD domain can be written as
\begin{equation}
  \label{eq_7}
  \begin{aligned}
  \mathbf{y}_{i}=&\sum\nolimits_{p=1}^{P_{i}}h_{i,p}\left(\mathbf{f}_{p}^{H}\mathbf{b}\left ( \theta _{i,p} \right )\right)\left (\mathbf{F}_{N}\otimes \mathbf{I}_{M}  \right )\mathbf{\Pi}^{l_{i,p}}\mathbf{\Delta}^{k_{i,p}}\\
  &\left (\mathbf{F}_{N}^{H}\otimes \mathbf{I}_{M}  \right )\mathbf{x}_{i}+\mathbf{n}_{i},
\end{aligned} 
\end{equation}
where $\mathbf{n}_{i}=\left (\mathbf{F}_{N}\otimes \mathbf{I}_{M}  \right )\mathbf{q}_{i}$ denotes the noise vector in the DD domain.

\vspace{-0.3cm}
\section{ Joint Channel Estimation, Signal Detection and Sensing Parameter Association } 

%In this section, we formulate the joint PACESD as a constrained bilinear recovery problem and leveraging the Bi-UAMP algorithm, develop a message passing algorithm to solve it.

\subsection {Problem Formulation}
We consider uplink transmission of vehicle $i$. Defining $\mathbf{G}_{i,p}=\left(\mathbf{f}_{p}^{H}\mathbf{b}\left ( \theta _{i,p} \right )\right)\left (\mathbf{F}_{N}\otimes \mathbf{I}_{M}  \right )\mathbf{\Pi}^{l_{i,p}}\mathbf{\Delta}^{k_{i,p}}\left (\mathbf{F}_{N}^{H}\otimes \mathbf{I}_{M}  \right )$$\in \mathbb{C}^{MN\times MN}$, \eqref{eq_7} can be rewritten as 
\begin{equation}
  \begin{aligned}
    \label{eq_8}
  \mathbf{y}_{i}=\sum\nolimits_{p=1}^{P_{i}}h_{i,p}\mathbf{G}_{i,p}\mathbf{x}_{i}+\mathbf{n}_{i}.
\end{aligned}
\end{equation}
We have the following remarks:  
\begin{itemize}
    \item According to the workflow of the vehicular network in Section II.A, the RSU possesses the sensing parameters from Steps 1 and 2, including time delays, arrival angles, and Doppler frequencies of the paths for all objects in the sensing area. These parameters can be used to construct $\mathbf{G}_{i,p}$ in \eqref{eq_8}, so that the RSU only needs to estimate $h_{i,p}, (p=1,\dots,P_i)$ and $\mathbf{x}_{i}$. So, the uplink transmission can be assisted by sensing.
    \item However, we note that the sensing parameters at the RSU are yet to be associated with the vehicles. Model \eqref{eq_8} is only available when the parameters have been associated with the vehicles.      
\end{itemize}

In this work, we propose a novel method to achieve joint PACESD.   
%In the following, we will use the Bi-UAMP algorithm on \eqref{eq_9} to perform parameter association, together with joint channel estimation and signal detection.
Let $P'=\sum_{i=1}^{K}P_{i}$, where $K$ is the number of targets/vehicles. We can construct the following model 
\begin{equation}
  \begin{aligned}
    \label{eq_9}
\mathbf{y}_{i}=\sum\nolimits_{p'=1}^{P'}h_{p'}{\mathbf{G}}_{p'}\mathbf{x}_{i}+\mathbf{n}_{i}.
\end{aligned}
\end{equation}
Regarding the model, we note the following:
\begin{itemize}
\item As the RSU does not know which sensing parameters belong to vehicle $i$,  ${\mathbf{G}}_{p'} (p'=1,\dots,P')$ is constructed using all sensing parameters available at the RSU.
\item We can see that $h_{p'}$ should be nonzero if the corresponding ${\mathbf{G}}_{p'}$ belongs to vehicle $i$, otherwise $h_{p'}=0$. Hence $\mathbf{h}_{i}=\left[h_{1},\dots,h_{P'}\right]^{T}$ is a sparse vector.
\item To achieve joint PACESD with model \eqref{eq_9}, we need to recover both $\mathbf{h}_{i}$ and  $\mathbf{x}_{i}$ based on $\mathbf{y}_{i}$  at the same time. This is a bilinear recovery problem. Also, we note that the constraints on the bilinear recovery, i.e., $\mathbf{h}_{i}$ is sparse and the entries of $\mathbf{x}_{i}$ are discrete-valued as they are transmitted symbols of vehicle $i$.  
\end{itemize}
Next, leveraging the Bi-UAMP algorithm \cite{ref13}, we develop a message passing algorithm as the core of the PACESD algorithm.
%where $\hat{\mathbf{G}}_{p'}$ is constructed from The estimates of sensing parameters for each target (i.e., $\hat{l}_{p'}$, $\hat{k}_{p'}$ and $\hat{\theta}_{p'}$) obtained in Step 2 and  assuming $\mathbf{f}_{p}=\mathbf{b}^{H}\left ( \hat{\theta} _{p'} \right )$.

%and $\mathbf{h}_{i}=\left[h_{1},\dots,h_{P'}\right]^{T}$. Because $P'>P_{i}$, only some elements in $\mathbf{h}_{i}$ are non-zero, which correspond to the source locations.

\vspace{-0.4cm}

\subsection{Probabilistic Representation and Message Passing Algorithm Design} 
Following the Bi-UAMP algorithm, we define ${\mathbf{G}}=\left [ {\mathbf{G}}_{1},\dots ,{\mathbf{G}}_{P'} \right ] \in \mathbb{C}^{MN\times MNP'}$, and rearrange the order of the columns of ${\mathbf{G}}$ to get ${\boldsymbol{\Psi}}=[\boldsymbol{\Psi}_{1},\dots ,\boldsymbol{\Psi}_{\dot{M}}]\in \mathbb{C}^{MN\times MNP'}$, so that \eqref{eq_9} can be rewritten as 
\begin{equation}
  \begin{aligned}
  \label{eq_10}
  \mathbf{y}_{i}={\boldsymbol{\Psi}}\mathbf{c}_{i}+\mathbf{n}_{i},
  \end{aligned}    
\end{equation}
with the auxiliary vector
\begin{equation}
  \begin{aligned}
  \label{eq_11}
  \mathbf{c}_{i}=\mathbf{x}_{i}\otimes\mathbf{h}_{i}=\left [ x_{1}\mathbf{h}_{i}^{T},\dots,x_{\dot{M}}\mathbf{h}_{i}^{T} \right ]^{T},
  \end{aligned}    
\end{equation}
where $\mathbf{c}_{i}= [ c_{1,1},\dots,c_{1,P'},\dots,c_{\dot{m} ,p'},\dots,c_{MN,P'}]^{T}=\left [ \mathbf{c}_{1}^{T},\dots,\mathbf{c}_{\dot{m}}^{T},\dots,\mathbf{c}_{\dot{M}}^{T}\right ]^{T}$ with $c_{\dot{m} ,p'}=x_{\dot{m}}h_{p'}$, $1\le\dot{m}\le \dot{M}= MN$.
\begin{algorithm}[tt]
  \caption{Bi-UAMP based algorithm for joint PACESD }\label{al_1}
  {\small{
  \begin{algorithmic}[1] 
\renewcommand{\algorithmicrequire}
{ \textbf{Define:}}
 \REQUIRE{%$\mathbf{r}_{i}=\mathbf{U}^{H}\mathbf{y}_{i}=\boldsymbol{\Phi}\mathbf{c}_{i}+\boldsymbol{\omega}_{i}$, where ${\mathbf{G}}=\mathbf{U}\mathbf{\Lambda}\mathbf{V}^{\mathrm{H}}$,$\boldsymbol{\Phi}=\mathbf{\Lambda}\mathbf{V}^{H}$, and $\mathbf{c}_{i}=\mathbf{h}_{i}\otimes\mathbf{x}_{i}$ with $\mathbf{h}_{i}=\left [ h_{1},\dots ,h_{P'} \right ]^{T}$ and $\mathbf{x}_{i}=\left [ x_{1},\dots ,x_{\dot{M}} \right ]^{T}$.\\
 $\mathbf{\Phi} =\left [\mathbf{\Phi}_{1},\dots,\mathbf{\Phi}_{\dot{M}}  \right ]$, $ \phi _{\dot{m}}=\left |  \mathbf{\Phi}_{\dot{m}}\right |^{2}\mathbf{1}_{P'}$, and $\mathbf{c}_{i}=\left [ \mathbf{c}_{1}^{T},\dots,\mathbf{c}_{\dot{M}}^{T}\right ]^{T}$, $\dot{m}\in[1,\dot{M}]$ and $p'\in[1,P']$.
 }
 {\renewcommand{\algorithmicrequire}{ \textbf{Initialization:}}
 \REQUIRE 
 $v_{x_{\dot{m}}}=1$, $v_{\mathbf{c}_{\dot{m}}}=1$, $\hat{\mathbf{c}}_{\dot{m}}=\mathbf{0}$, $\mathbf{z}=\mathbf{0}$, $\mathbf{s}=\mathbf{0}$, $\hat{\beta}=1$.}
 \renewcommand{\algorithmicrequire}{ \textbf{Repeat:}}
 \REQUIRE{} 
  \que{} \\  \textbf{\quad Param. assoc.} and \textbf{chan. est.}: Lines 1-15
  \STATE $\mathbf{v}_{\mathbf{p}}=\sum_{\dot{m}}\phi _{\dot{m}}v_{\mathbf{c}_{\dot{m}}}$
  \STATE $\mathbf{p}=\sum_{\dot{m}}\mathbf{\Phi}_{\dot{m}}\hat{\mathbf{c}}_{\dot{m}}-\mathbf{v}_{\mathbf{p}}\cdot\mathbf{z}$
  \STATE $\mathbf{v}_{\boldsymbol{\zeta}}=\mathbf{v}_{\mathbf{p}}./( \mathbf{1}+\hat{\beta}\mathbf{v}_{\mathbf{p}})$
  \STATE $\hat{\boldsymbol{\zeta}}=( \hat{\beta}\mathbf{v}_{\mathbf{p}}\cdot\mathbf{r}_{i}+\mathbf{p})./( \mathbf{1}+\hat{\beta}\mathbf{v}_{\mathbf{p}})$
  \STATE $\hat{\beta}=\dot{M}/(\| \mathbf{r}-\hat{\boldsymbol{\zeta}}\|^{2}+\mathbf{1}^{T}\mathbf{v}_{\boldsymbol{\zeta}})$
  \STATE $\mathbf{v}_{\mathbf{z}}=\mathbf{1}./( \mathbf{v}_{\mathbf{p}}+\hat{\beta}^{-1}\mathbf{1}_{P'})$
  \STATE $\mathbf{z}=\mathbf{v}_{\mathbf{z}}\cdot( \mathbf{r}-\mathbf{p})$
  \STATE $\forall \dot{m}:v_{\mathbf{q}_{\dot{m}}}=1/\langle | \mathbf{\Phi}_{\dot{m}}^{H} |^{2} \mathbf{v}_{\mathbf{s}}\rangle$
  \STATE $\forall \dot{m}:\mathbf{q}_{\dot{m}}=\hat{\mathbf{c}}_{\dot{m}}+v_{\mathbf{q}_{\dot{m}}}\mathbf{\Phi}_{\dot{m}}^{H}\mathbf{z}$
  %\\ \textbf{\% Signal detection}: Lines 10-15
%  \\ \quad \yx{(In the case of $x_{\ddot{m}},\ddot{m}=(128:128:\dot{M})$ is known, set $\hat{x}_{\ddot{m}}=x_{\ddot{m}}$ and $v_{x_{\ddot{m}}}=0$.)}
  \STATE $\forall \dot{m}:\vec{\mathbf{v}}_{\mathbf{h}_{\dot{m}}}=\mathbf{1}_{P'}v_{\mathbf{q}_{\dot{m}}}./(|\hat{x}_{\dot{m}}|^2+\mathbf{v}_{x_{\dot{m}}})$
  \STATE $\forall \dot{m}:\vec{\mathbf{h}}_{\dot{m}}=\mathbf{q}_{\dot{m}}\cdot\hat{x}_{\dot{m}}^{*}./(|\hat{x}_{\dot{m}}|^2+\mathbf{v}_{x_{\dot{m}}})$
  \STATE $\vec{\mathbf{v}}_{\mathbf{h}_{i}}=\mathbf{1}_{P'}./(\sum_{\dot{m}}\mathbf{1}_{P'}./\vec{\mathbf{v}}_{\mathbf{h}_{\dot{m}}})$
  \STATE $\vec{\mathbf{h}_{i}}=\vec{\mathbf{v}}_{\mathbf{h}_{i}}\cdot\sum_{\dot{m}}(\vec{\mathbf{h}}_{\dot{m}}./\vec{\mathbf{v}}_{\mathbf{h}_{\dot{m}}})$
  \STATE compute (18) and (19)
  \STATE $\mathbf{v}_{\mathbf{h}_{i}}= <[ v_{{h}_{1}},\dots,v_{{h}_{P'}}]>\mathbf{1}_{P'}$, $\hat{\mathbf{h}}_{i}=[ \hat{h}_{1},\dots,\hat{h}_{P'}]^{T } $
  \\ \textbf{\quad Sig. det.}: Lines 16-21
  \STATE $\forall \dot{m}:\vec{\mathbf{v}}_{\mathbf{x}_{\dot{m}}}=v_{\mathbf{q}_{\dot{m}}}\mathbf{1}_{P'}./(|\hat{\mathbf{h}}|^{2}+\mathbf{v}_{\mathbf{h}})$
  \STATE $\forall \dot{m}:\vec{\mathbf{x}}_{\dot{m}}=\mathbf{q}_{\dot{m}}\hat{\mathbf{h}}^{*}/(|\hat{\mathbf{h}}|^{2}+\mathbf{v}_{\mathbf{h}})$
  \STATE $\forall \dot{m}:\vec{v}_{x_{\dot{m}}}=(\mathbf{1}_{P'}^{T}(\mathbf{1}_{P'}./\vec{\mathbf{v}}_{\mathbf{x}_{\dot{m}}}))^{-1}$
  \STATE $\forall \dot{m}:\vec{x}_{\dot{m}}=\vec{v}_{x_{\dot{m}}}\mathbf{1}_{P'}^{T}(\vec{\mathbf{x}}_{\dot{m}}./\vec{\mathbf{v}}_{\mathbf{x}_{\dot{m}}})$
  \STATE compute (25) and (26) 
  \STATE $\mathbf{v}_{\mathbf{x}_{i}}=\langle [ v_{{x}_{1}},\dots,v_{{x}_{\dot{M}}}]  \rangle \mathbf{1}_{\dot{M}}$, $\hat{\mathbf{x}}_{i}=[ \hat{x}_{1},\dots,\hat{x}_{\dot{M}}]^{T } $
% \\ \quad \yx{(In the case of $x_{\ddot{m}},\ddot{m}=(128:128:\dot{M})$ is known, set $\hat{x}_{\ddot{m}}=x_{\ddot{m}}$ and $v_{x_{\ddot{m}}}=0$.)}
  \STATE $\forall \dot{m}:\overleftarrow{\mathbf{v}}_{\mathbf{x}_{\dot{m}}}=(v_{x_{\dot{m}}}\vec{\mathbf{v}}_{\mathbf{x}_{\dot{m}}})./(\vec{\mathbf{v}}_{\mathbf{x}_{\dot{m}}}-v_{x_{\dot{m}}}\mathbf{1}_{P'})$
  \STATE $\forall \dot{m}:\overleftarrow{\mathbf{x}}_{\dot{m}}= (\hat{x}_{\dot{m}}\vec{\mathbf{v}}_{\mathbf{x}_{\dot{m}}}- v_{x_{\dot{m}}}\vec{\mathbf{x}}_{\dot{m}})./(\vec{\mathbf{v}}_{\mathbf{x}_{\dot{m}}}-v_{x_{\dot{m}}}\mathbf{1}_{P'})$
  \STATE $\forall \dot{m}:\overleftarrow{\mathbf{v}}_{\mathbf{h}_{\dot{m}}}=(\mathbf{1}./\mathbf{v}_{\mathbf{h}_{i}}-\mathbf{1}./\vec{\mathbf{v}}_{\mathbf{h}_{\dot{m}}})^{-1}$
  \STATE $\forall \dot{m}:\overleftarrow{\mathbf{h}}_{\dot{m}}=\overleftarrow{\mathbf{v}}_{\mathbf{h}_{\dot{m}}}\cdot(\hat{\mathbf{h}}_{i}./\mathbf{v}_{\mathbf{h}_{i}}-\vec{\mathbf{h}}_{\dot{m}}./\vec{\mathbf{v}}_{\mathbf{h}_{\dot{m}}})$
  \STATE $\forall \dot{m}:\overleftarrow{\mathbf{c}}_{\dot{m}}=\overleftarrow{\mathbf{x}}_{\dot{m}}\cdot\overleftarrow{\mathbf{h}}_{\dot{m}}$
  \STATE $\forall \dot{m}:\overleftarrow{\mathbf{v}}_{\mathbf{c}_{\dot{m}}}=|\overleftarrow{\mathbf{x}}_{\dot{m}}|^{2}\cdot\overleftarrow{\mathbf{v}}_{\mathbf{h}_{\dot{m}}}+\overleftarrow{\mathbf{v}}_{\mathbf{x}_{\dot{m}}}\cdot|\overleftarrow{\mathbf{h}}_{\dot{m}}|^{2}+\overleftarrow{\mathbf{v}}_{\mathbf{x}_{\dot{m}}}\cdot\overleftarrow{\mathbf{v}}_{\mathbf{h}_{\dot{m}}}$
  \STATE $\forall \dot{m}:\mathbf{v}_{\mathbf{c}_{\dot{m}}}=(1/v_{\mathbf{q}_{\dot{m}}}\mathbf{1}_{P'}+\mathbf{1}./\overleftarrow{\mathbf{v}}_{\mathbf{c}_{\dot{m}}})^{\mathrm{-1}}$
  \STATE $\forall \dot{m}:\hat{\mathbf{c}}_{\dot{m}}=\mathbf{v}_{\mathbf{c}_{\dot{m}}}\cdot(1/v_{\mathbf{q}_{\dot{m}}}\mathbf{q}_{\dot{m}}+\overleftarrow{\mathbf{c}}_{\dot{m}}./\overleftarrow{\mathbf{v}}_{\mathbf{c}_{\dot{m}}})$
  \STATE $\forall \dot{m}:v_{\mathbf{c}_{\dot{m}}}=\langle \mathbf{v}_{\mathbf{c}_{\dot{m}}} \rangle$
  \renewcommand{\algorithmicrequire}{ \textbf{Until terminated}}
  \REQUIRE{}
  
  \end{algorithmic}}}
  \end{algorithm}
  
To enable the use of UAMP \cite{ref16,ref17}, 
we perform singular value decomposition (SVD) of the matrix $\boldsymbol{\Psi}$ , i.e., $\boldsymbol{\Psi}=\mathbf{U}\mathbf{\Lambda}\mathbf{V}^{H}$, and carry out a unitary transformation with $\mathbf{U}^{H}$ on \eqref{eq_10}, yielding
\begin{equation}
    \begin{aligned}
    \label{eq_12}
    \mathbf{r}_{i}=\boldsymbol{\Phi}\mathbf{c}_{i}+\boldsymbol{\omega}_{i},
    \end{aligned}    
\end{equation}
where $\mathbf{r}_{i}=\mathbf{U}^{H}\mathbf{y}_{i}$,$\boldsymbol{\Phi}=\mathbf{\Lambda}\mathbf{V}^{\mathrm{H}}$ and $\boldsymbol{\omega}_{i}=\mathbf{U}^{H}\mathbf{n}_{i}$ is still AWGN with the precision $\beta$. We assume that $\beta$ is unknown, which needs to be estimated as well. 
With \eqref{eq_11} and \eqref{eq_12}, we aim to recover $\mathbf{h}_{i}$ and $\mathbf{x}_{i}$. Inspired by the sparse Bayesian learning technique, we use sparsity-promoting hierarchical Gaussian-Gamma distribution for the entries in $\mathbf{h}_{i}$, i.e., 
\begin{equation}
  \label{eq_13}
    p(\mathbf{h}_{i}|\boldsymbol{\gamma})=\prod\nolimits_{p'} p(h_{p'}|\gamma_{p'})= \prod\nolimits_{p'} \mathcal{N} (h_{p'}; 0, \gamma_{p'}^{-1}), 
\end{equation} 
where the precisions $\gamma_{p'}$ are Gamma distributed, i.e.,
\begin{equation}
  \label{eq_14}
    p(\gamma_{p'})= \text{Ga}\left(\gamma_{p'};\epsilon,\eta\right),\forall p'
\end{equation}
with $\epsilon$ and $\eta$ being the shape and rate parameters. 
The prior of $\mathbf{x}_{i}$ can be expressed as
\begin{equation}
  \label{eq_15}
    p(\mathbf{x}_i)= {1}/{\left|\mathcal{A}\right|} \prod\nolimits_{\dot{m}} \sum\nolimits_{a=1}^{\left|\mathcal{A}\right|} \delta(x_{\dot{m}}-\alpha_{a}),
\end{equation}
where $\mathcal{A}$ denotes the alphabet of the symbols in DD domain, i.e., $\alpha_{a}\in\mathcal{A}=\left\{\alpha_{1},\dots,\alpha_{\left|\mathcal{A}\right|}\right\}$.% with $\left|\mathcal{A}\right|$ being the cardinality of $\mathcal{A}$.

\par
Define an auxiliary variable $\boldsymbol{\zeta }=\boldsymbol{\Phi}\mathbf{c}_{i}$. The joint conditional distribution of the unknown variables can be expressed as
%\begin{eqnarray}    &&\!\!\!\!\!\!\!\!\!\!\!\!\!\!\!p(\mathbf{h}_{i},\mathbf{x}_{i},\mathbf{c}_{i},\boldsymbol{\zeta},\beta|\mathbf{r}) \propto p(\mathbf{r}|\boldsymbol{\zeta },\beta) \nonumber \\
%&&p(\boldsymbol{\zeta}|\mathbf{c}_{i}) p(\mathbf{c}_{i}|\mathbf{h}_{i},\mathbf{x}_{i})p(\mathbf{h}_{i})p(\mathbf{x}_{i})p(\beta),
%\end{eqnarray}
{\begin{equation}
  \begin{aligned}
  p(\mathbf{h}_{i},\mathbf{x}_{i},\mathbf{c}_{i},\boldsymbol{\zeta},\beta|\mathbf{r}) \propto& \;p(\mathbf{r}|\boldsymbol{\zeta },\beta)p(\boldsymbol{\zeta}|\mathbf{c}_{i})\\
  &\times p(\mathbf{c}_{i}|\mathbf{h}_{i},\mathbf{x}_{i})p(\mathbf{h}_{i})p(\mathbf{x}_{i})p(\beta),
  \end{aligned}
  \end{equation}}
where $p(\mathbf{r}|\boldsymbol{\zeta},\beta)=\mathcal{N}\left(\boldsymbol{\zeta};\mathbf{r},\beta^{-1}\mathbf{I}\right)$, $p(\boldsymbol{\zeta}|\mathbf{c}_{i})=\delta\left(\boldsymbol{\zeta}-\boldsymbol{\Phi}\mathbf{c}_{i}\right)$, $p(\mathbf{c}_{i}|\mathbf{h}_{i},\mathbf{x}_{i})=\delta\left(\mathbf{c}_{i}-\mathbf{x}_{i}\otimes\mathbf{h}_{i}\right)$ and $p(\beta)\propto\beta^{-1}$.
Following the Bi-UAMP algorithm, we can compute the (approximate) \textit{a posteriori} distributions $p(x_{\dot{m}}|\mathbf{r}_i)$ and $p(h_{p'}|\mathbf{r}_i)$, so that their estimates in terms of the \textit{a posteriori} means can be obtained.
\begin{figure}[tt]
  \setlength{\abovecaptionskip}{0.cm}
  \setlength{\belowcaptionskip}{-0.cm}
    \centering
    \vspace{-0.3cm}
    \includegraphics[width=2in]{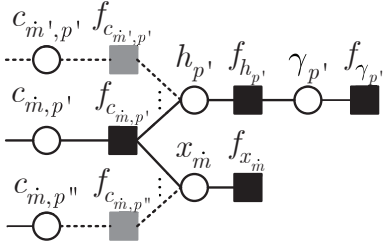}
    \caption{Part of the factor graph representation.}
    \label{fig_2}
    \vspace{-0.5cm}
  \end{figure}
\par
It is noted that Bi-UAMP in \cite{ref13} is algorithmic framework, which does not specify the priors of the variables to be recovered. %i.e., $\mathbf{h}_{i}$ and $\mathbf{x}_{i}$ considered in this work. 
Hence, we need to derive the concrete message updating rules related to the priors of $\mathbf{h}_{i}$ and $\mathbf{x}_{i}$. The relevant part of the factor graph is shown in Fig. \ref{fig_2}.  
The developed algorithm is summarized in Algorithm \ref{al_1}, with major steps detailed below. 

According to the derivation of Bi-UAMP, running Lines 1-13 produces a mean vector $\vec{\mathbf{h}}_{i}$ and variance vector $\vec{\mathbf{v}}_{\mathbf{h}_{i}}$. The message from variable node $h_{p'}$ to function node $f_{h_{p'}}$ in Fig. \ref{fig_2} follows a Gaussian distribution, i.e., $\mathcal{M}_{h_{p'}\rightarrow f_{h_{p'}}}(h_{p'})= \mathcal{N}(h_{p'}; \vec{h}_{p'}, \vec{v}_{h_{\dot{m}}})$, where $\vec{h}_{p'}$ and $\vec{v}_{h_{p'}}$ are the $p'$-th element of $\vec{\mathbf{h}}_{i}$ and $\vec{\mathbf{v}}_{\mathbf{h}_{i}}$, respectively.
Combining it with the message $\mathcal{M}_{f_{h_{p'}}\rightarrow h_{p'}}(h_{p'})=\mathcal{N}(h_{p'};0,\hat{\gamma}_{p'}^{-1})$ (i.e., prior \eqref{eq_13}), we obtain the belief of $h_{p'}$
\begin{equation}
  \begin{aligned}
    b(h_{p'})&\propto\mathcal{M}_{h_{p'}\rightarrow f_{h_{p'}}}(h_{p'})\mathcal{M}_{f_{h_{p'}}\rightarrow h_{p'}}(h_{p'})\\
    &=\mathcal{N}(h_{p'}\rev{;}\hat{h}_{p'},v_{h_{p'}}),
  \end{aligned}
\end{equation}
where 
\begin{equation}
  \begin{aligned}
    \hat{h}_{p'}&=\vec{h}_{p'}(1/\vec{v}_{h_{p'}}+\hat{\gamma}_{p'})^{-1},   
  \end{aligned}
\end{equation}
\begin{equation}
  \begin{aligned}
    v_{h_{p'}}&=(1/\vec{v}_{h_{p'}}+\hat{\gamma}_{p'})^{-1},
  \end{aligned}
\end{equation}
where the computation of $\hat{\gamma}_{p'}$ is given in \eqref{precision}. 
%Performing the average operations on $v_{h_{p'}}$ in (23) and arranging (22) in a vector form lead to Line 21 of the algorithm.
%Since the function node $f_{h_{p'}}$ connects the variable node $\gamma_{p'}$, 
According to the mean field rule, message $\mathcal{M}_{f_{h_{p'}}\rightarrow\gamma_{p'}}(\gamma_{p'})$ can be expressed as 
\begin{equation}
  \begin{aligned}
   \mathcal{M}_{f_{h_{p'}}\rightarrow\gamma_{p'}}(\gamma_{p'})&\!\propto\!\text{exp}\{<\text{log}f_{h_{p'}}(h_{p'}|0,\gamma_{p'}^{-1})>_{b(h_{p'})}\}\\
    &\!\propto\!\sqrt{\gamma_{p'}}\text{exp}\{-\frac{\gamma_{p'}}{2}(|\hat{h}_{p'}|^{2}+v_{h_{p'}})\}.
  \end{aligned}
\end{equation}
Combining it with the message $\mathcal{M}_{f_{\gamma_{p'}}\rightarrow \gamma_{p'}}(\gamma_{p'})\propto \gamma_{p'}^{\epsilon-1}\text{exp}\{-\eta\gamma_{p'}\}$ (i.e., prior \eqref{eq_14}), we can get the belief of $\gamma_{p'}$
\begin{equation}
  \begin{aligned}
    b(\gamma_{p'})&\propto\mathcal{M}_{f_{\gamma_{p'}}\rightarrow \gamma_{p'}}(\gamma_{p'})\mathcal{M}_{f_{h_{p'}}\rightarrow\gamma_{p'}}(\gamma_{p'})\\
    &\propto\gamma_{p'}^{\epsilon-\frac{1}{2}}\text{exp}\{-\frac{\gamma_{p'}}{2}(|\hat{h}_{p'}|^{2}+v_{h_{p'}}+2\eta)\}
  \end{aligned}
\end{equation}
and 
\begin{equation} \label{precision}
  \begin{aligned}
    \hat{\gamma}_{p'}=<\gamma_{p'}>_{b(\gamma_{p'})}=(2\epsilon+1)/(|\hat{h}_{p'}|^{2}+v_{h_{p'}}+2\eta),
  \end{aligned}
\end{equation}
where $\eta$ is set to 0. For the shape parameter $\epsilon$, we use the automatic tuning rule proposed in \cite{ref16}
\begin{equation}
  \epsilon=\frac{1}{2} \sqrt{\mathrm{log}( \frac{1}{P'}\sum\nolimits_{p'=1}^{P'}\hat{\gamma}_{p'} )-\frac{1}{P'}\sum\nolimits_{p'=1}^{P'}\mathrm{log}\hat{\gamma}_{p'}}.
\end{equation}
The above results lead to Line 14 of the algorithm. Following Bi-UAMP, performing the average operations on $v_{h_{p'}}$ in (18) and arranging (17) in a vector form lead to Line 15 of the algorithm.
\par
According to Bi-UAMP, Lines 16-19 generate  $\vec{x}_{\dot{m}}$ and $\vec{v}_{x_{\dot{m}}}$. The message from variable node $x_{\dot{m}}$ to function node $f_{x_{\dot{m}}}$ in Fig. \ref{fig_2} also follows a Gaussian distribution, i.e., $\mathcal{M}_{x_{\dot{m}}\rightarrow f_{x_{\dot{m}}}}(x_{\dot{m}})= \mathcal{N}(x_{\dot{m}}; \vec{x}_{\dot{m}}, \vec{v}_{x_{\dot{m}}})$, where $\vec{x}_{\dot{m}}$ and $\vec{v}_{x_{\dot{m}}}$ are the $\dot{m}$-th element of $\vec{\mathbf{x}}_{i}$ and $\vec{\mathbf{v}}_{\mathbf{x}_{i}}$, respectively.
Hence, we have the following pseudo observation model   %In particular, Line 14 is about the computation of the a posterior means and variances of the symbols with the following pseudo observation model
\begin{equation}
  \label{eq_24}
    \vec{x}_{\dot{m}}=x_{\dot{m}}+\varpi_{\dot{m}},\forall \dot{m}
\end{equation}
{where 
$\varpi_{\dot{m}}$ denotes a model Gaussian noise with mean 0 and variance $\vec{v}_{x_{\dot{m}}}$. According to Bi-UAMP, with the prior \eqref{eq_15} and model \eqref{eq_24}, 
we compute the \textit{a posteriori} mean and variance of $x_{\dot{m}}$, which are given as }
\begin{equation}
    \begin{aligned}
    \label{eq_25}
    \hat{x}_{\dot{m}}=\sum\nolimits_{a=1}^{\left|\mathcal{A}\right|}\alpha_{a}\varrho_{\dot{m},a}, \forall \dot{m}
    \end{aligned}    
\end{equation}
and
\begin{equation}
    \begin{aligned}
    \label{eq_26}
    v_{x_{\dot{m}}}=\sum\nolimits_{a=1}^{\left|\mathcal{A}\right|}\varrho_{\dot{m},a}|\alpha_{a}-\hat{x}_{\dot{m}}|^{2}, \forall \dot{m}
    \end{aligned}    
\end{equation}
where $\varrho_{\dot{m},a}=\tau_{\dot{m},a}/\sum_{a=1}^{\left|\mathcal{A}\right|}\tau_{\dot{m},a}$ and $\tau_{\dot{m},a}=\mathrm{exp}(-\vec{v}_{x_{\dot{m}}}^{-1}|\alpha_{a}-\vec{x}_{\dot{m}}|^{2})$.
%\begin{equation}
%\tau_{\dot{m},a}=\mathrm{exp}(-\vec{v}_{x_{\dot{m}}}^{-1}|\alpha_{a}-\vec{x}_{\dot{m}}|^{2}).
%\end{equation}
In the last iteration, hard decisions are made to the symbols based on the \textit{a posteriori} means $\{\hat{x}_{\dot{m}}\}$.
The above operations correspond to Line 20. Lines 21-30 are obtained following Bi-UAMP. 
\par
It is noted that there is an inherent ambiguity problem with a bilinear problem. To mitigate the ambiguity, we assume a very small fraction ($1/128$) of the symbols in $\mathbf{x}_i$ is known at the RSU. This only leads to an overhead about 0.78\%. %can be ignored. 
Algorithm \ref{al_1} requires an SVD for pre-processing, and the complexity is $\mathcal{O}(NM^{2}P'\text{log}(MP'))$ with a modern SVD algorithm \cite{ref18}. 
The proposed algorithm does not involve matrix inversion, and the complexity of each iteration is dominated by the matrix-vector products, which is $\mathcal{O}(\dot{M}P')+\mathcal{O}(|\mathcal{A}|)$ per symbol.

\vspace{-0.3cm}

\section{Simulation Results}
\vspace{-0.1cm}
In the simulation, we set $N_{\mathrm{BS}}=128$, $M=128$, $N=32$, carrier frequency $f_c=4$ GHz and subcarrier spacing $\Delta f=15$ kHz.
We assume the number of targets/vehicles $K=3$, the number of channel paths between each target/vehicle and RSU is 6, the maximum Doppler index $k_{\mathrm{max}}=6$, and the maximum delay index $l_{\mathrm{max}}=6$.
QPSK modulation is used.
%\yx{We compare the proposed detector with the MMSE detector, where perfect CSI is used.}
%We compare the proposed detector with the ML detector with superimposed pilot in \cite{ref12}, where zero forcing (ZF) channel estimation is used, and the ratio of pilot to data power is denoted by $R_p$.

\begin{figure}[tt]
  \setlength{\abovecaptionskip}{0.cm}
  \setlength{\belowcaptionskip}{-.cm}
    \centering
    \vspace{-0.3cm}
    \includegraphics[width=3.25in]{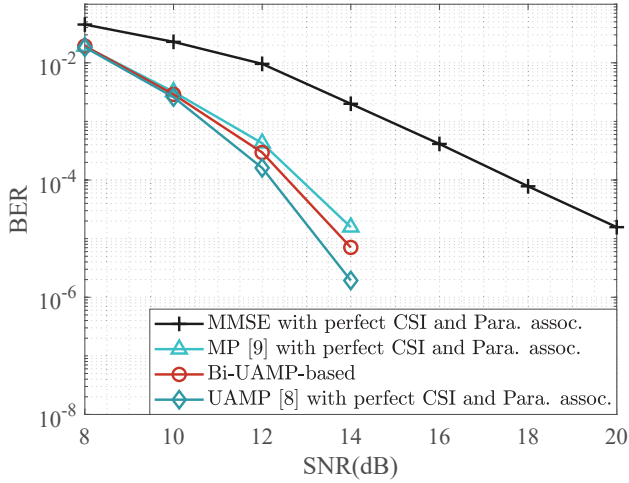}
    \caption{BER performance of sensing-aided uplink transmission.}
    \label{fig_3}
  \end{figure}

Fig. \ref{fig_3} shows the BER performance of the proposed algorithm. To the best of our knowledge, parameter association has not been considered in the literature. To facilitate the comparison, we assume perfect CSI and parameter association for the coventional MMSE detector, the MP based detector in \cite{ref9}, and the UAMP-based detector \cite{ref8} (serving as a BER lower bound).
It can be seen from Fig. \ref{fig_3} that our proposed algorithm achieves performance close to the bound, %incurring only \yx{0.5dB degradation at BER of $4.7\times 10^{-5}$.  
and outperforms the other two detectors.
%For the method in [12], a small pilot power results in inaccurate channel estimation, thus poor detection performance, while a high pilot power will result in smaller power for the data signal, leading to performance loss as well.  
%Compared to the scheme with guard symbols based pilot \cite{ref10}, the proposed scheme can improve the spectral efficiency by $\frac{\left ( 2l_{max}+1 \right )\left ( 4k_{max}+1 \right )}{MN+L_{CP}}\approx 17.67\%$ with close performance, where $L_{CP}=l_{max}$ denotes the length of cyclic prefix.
  
\begin{figure}[!ht]
\setlength{\abovecaptionskip}{0.cm}
\setlength{\belowcaptionskip}{-0.cm}
  \centering
  \vspace{-0.3cm}
  \includegraphics[width=3.25in]{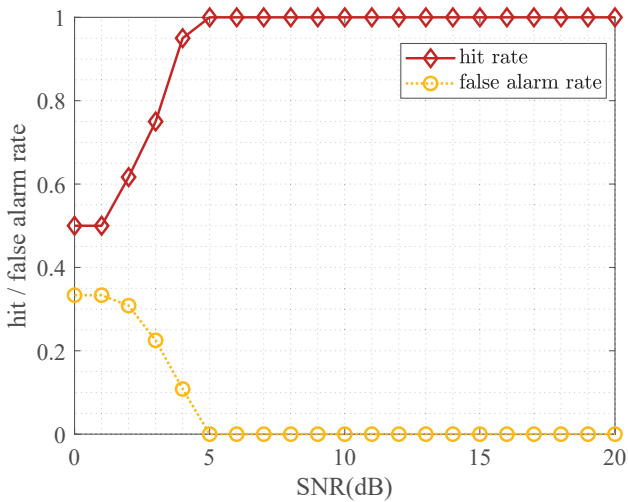}
  \caption{Performance of parameter association.}
  \label{fig_4}
  \vspace{-0.4cm} 
\end{figure}
\begin{figure}[tt]
\setlength{\abovecaptionskip}{0.cm}
\setlength{\belowcaptionskip}{-0.cm}
  \centering
  \vspace{-0.3cm}
  \includegraphics[width=3.25in]{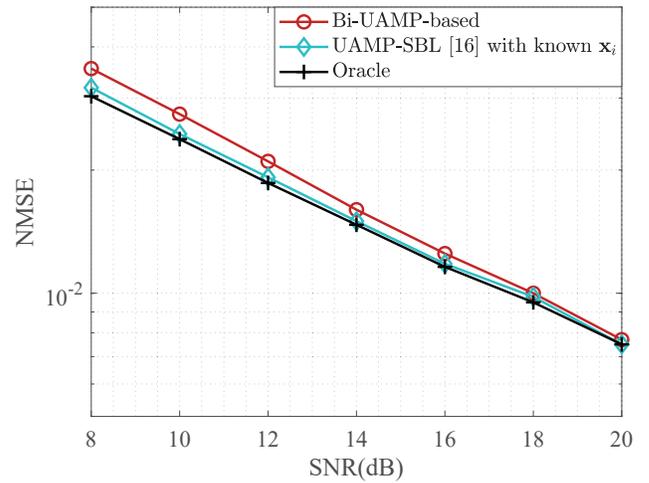}
  \caption{NMSE performance of channel estimation.}
  \label{fig_5}
  \vspace{-0.4cm}
\end{figure}

\par

\par
Fig. \ref{fig_4} shows the parameter association performance in terms of hit rate (i.e., the rate that all non-zero elements in $\mathbf{h}_{i}$ are correctly detected) and false alarm rate (i.e., the rate that zero elements in $\mathbf{h}_{i}$ are detected as non-zero elements).
It can be seen that when the SNR is larger than 5 dB, the proposed algorithm can almost achieve a 100\% hit rate, and zero false alarm rate for parameter association.
Fig. 5 compares the normalized mean squared error (NMSE) performance for channel estimation. An oracle performance bound is also included, which is obtained by performing MMSE estimation with perfect parameter association and known transmitted symbols $\mathbf{x}_i$.
It can be seen that the performance of our proposed Algorithm approaches that of the UAMP-SBL \cite{ref16} with known transmitted symbols $\mathbf{x}_i$, and also the oracle bound with the increase of SNR. 

\vspace{-0.2cm}
\section{Conclusion}
\vspace{-0.1cm}
In this paper, we have proposed a novel scheme for sensing-assisted uplink transmission in ISAC OTFS vehicle networks. To enable to exploit the acquired sensing parameters, joint PACESD is formulated as a bilinear recovery problem, which is solved by developing a message passing algorithm. Simulation results show the excellent performance of the scheme with the proposed algorithm.

\end{document}